\documentclass[10pt,letter]{rogics}
\usepackage{times,amssymb,amsmath,graphicx}
\usepackage[latin1]{inputenc}
\usepackage[OT1]{fontenc}

\title{\vspace*{-1cm}Pseudo-random Sequences Generated by Cellular Automata}
\author{\vspace*{-.3cm}Bruno Martin \and Patrick Solé\thanks{This work was
    supported by the french ANR program NUGET.}}
\institute{I3S, Université de Nice--Sophia Antipolis, CNRS,\\
2000 route des Lucioles, BP 121,
 F-06903 Sophia Antipolis Cedex \\
\email{\{Bruno.Martin$\mid$ Patrick.Sole\}@unice.fr}}

\begin{document}
\maketitle
\begin{abstract}
  \emph{\indent Generation of pseudo random sequences by cellular
    automata, as well as by hybrid cellular automata is surveyed.  An
    application to the fast evaluation and FPGA implementation of some
    classes of boolean functions is sketched out.}
\end{abstract}

\section*{Introduction}
Cellular Automata (CA) is a popular model of finite state machine with
some pretention to generality and universality.  Pseudo Random
Sequences (PRS) on the other hand, have a long history of applications
to computational (Monte Carlo sampling, numerical simulation) and
comunications problems (coding theory, streamciphers). In that context
the popular model is the Linear Feedback Shift Register (LFSR),
another model of linear finite state machine.

In the present work we survey the known attempts to generate PRS by
CA. We give an account of the synthesis of LFSR by arrays of variable
CA (known as hybrid CA or HCA). We sketch an application to the
evaluation of boolean functions in $n$ variables which are related to
cyclic codes of length $2^n-1.$ This is aimed at VLSI implementation,
especially by programmable arrays.

The material is organized as follows. Section~\ref{sec:notdef}
collects definitions and basic notions on PRS, CA and
HCA. Section~\ref{sec:synth} reviews the synthesis theory of LFSR by
HCA. Section~\ref{sec:classical-PRS} surveys the generation of PRS by
elementary CA. Section~\ref{sec:PRS-HCA} surveys the generation of
PRS by HCA.  Section~\ref{sec:appli} contains the application of
synthesis theory to boolean functions evaluation.

\section{Notations and definitions}\label{sec:notdef}

\subsection{(Pseudo-)randomness}

This section recalls the classical definitions of
pseudo-randomness. We first give an intuitive statement which gives
the difference between real randomness and pseudo-randomness. We then
introduce more formal definitions of pseudo-randomness.

\medskip

In~\cite{Wol:newkind}, Wolfram describes three mechanisms
responsible for random behavior in systems: (1) {\it Randomness from
  physics} like brownian motion; (2) {\it Randomness from the initial
  conditions} which is studied by chaos theory; and (3) {\it
  Randomness by design}, also called pseudo-randomness used in
pseudo-random sequences generators. Many algorithms generate
pseudo-random sequences. The behavior of the system is fully
determined by knowing the seed and the algorithm used. They are
quicker methods than getting "true" randomness from the environment,
inaccessible for computers.

The applications of randomness have led to many different methods for
generating random data. These methods may vary as to how unpredictable
or statistically random they are, and how quickly they can generate
random sequences.  Before the advent of computational PRS, generating large amount of sufficiently random numbers
(important in statistics and physical experimentation) required a lot
of work. Results would sometimes be collected and distributed as
random number tables or even CD iso-images.

\medskip

More formally, a pseudo-random sequence (PRS for short) can be defined
as:
\begin{definition}
  A sequence is {\em pseudo-random} if it cannot be
distinguished from a truly random sequence by any efficient
(polynomial time) procedure or circuit.
\end{definition}

\begin{theorem}[\cite{blum-micali84}]\label{th:blum}
  A sequence is {\em pseudo-random} iff it is next-bit
unpredictable.
\end{theorem}
Theorem~\ref{th:blum} claims that for pseudo-random sequences, even if
we know all the history, we don't have any
information on the next bit. Theorem~\ref{th:blum} was proved
equivalent to:
\begin{theorem}[\cite{yao-infoth}]\label{th:yao}
  A PRS generator $G$ passes Yao's test if, for any
  family of circuits $F$ with a polynomial number of gates for
  computing a statistical test, $G$ passes $F$.
\end{theorem}

\subsection{Cellular automata}
\label{sec:defs}

\begin{figure}[ht]
  \centering
  \includegraphics[scale=.85]{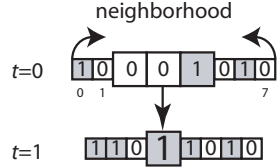}
  \caption{{\sf Transition of a cell (rule 30); cyclic CA.}}
  \label{fig:UneTrans}
\end{figure}

In this section, we recall several definitions of cellular
automata (CA). We focus on \emph{elementary} cellular
automata rules which restrict the set of
the states to be $\mathbb{F}_2$.  A {\em cellular automaton} is
generally a bi-infinite array of identical cells which evolve
synchronously and in parallel according to a local transition
function. The cells can only communicate with their nearest
neighbors. Here, we will concentrate on two finite restrictions of CA:
\begin{itemize}
\item cyclic:  a ring of $N$ cells indexed by
  $\mathbb{Z}_N$.
\item null boundary:  an array of $N$ cells
  in which both extremal cells are fed with zeroes.
\end{itemize}
All the cells are finite state machines with a finite number
of states and a transition function which gives the new state of a
cell according to its current state and the current states of its
nearest neighbors.
\begin{definition}\label{def:ca}
  A {\em cellular automaton} is a finite array of
  cells. Each cell is a finite state machine $C=(Q,f)$ where $Q$ is a
  finite set of states and $f$ a mapping $f:Q^3\rightarrow Q$.
\end{definition}
The mapping $f$, called {\em local transition function}, has the
following meaning: the state of cell $i$ at time $t+1$ (denoted by
$x_i^{t+1}$) depends upon the state of cells $i-1$, $i$ and $i+1$ at
time $t$ (the \emph{neighborhood} of cell $i$ of radius 1).
Fig.~\ref{fig:UneTrans} illustrates one transition of a cellular
automaton with 8 cells. The following equality
rules the dynamics of the cellular automaton:
\begin{equation}
  \label{eq:CArule}
  x_i^{t+1}=f(x_{i-1}^t,x_{i}^t,x_{i+1}^t)
\end{equation}
\vspace*{-.5cm}

\begin{figure}[ht]
  \centering
  \includegraphics[scale=.75]{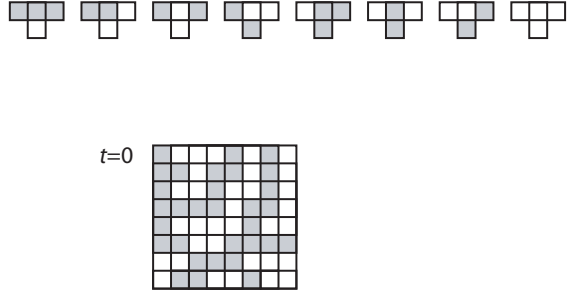}
  \caption{\sf{Evolution of CA30 on a ring with $N=8$ cells.}}
  \label{fig:CA30}
\end{figure}

\vspace*{-.5cm}
For a fixed $t$, the sequence of all the values $x_i^t$ for
$i\in{\mathbb Z}_N$, is the {\em configuration} at time $t$.  It is a
mapping $c$ which assigns a state of $Q$ to each cell of the cellular
automaton. The sequence of configurations as pictured by
Fig.~\ref{fig:CA30} is called a \emph{time-space}
diagram. Fig.~\ref{fig:CA30} depicts the evolution of a ring with
$N=8$ cells. On the top of Fig.~\ref{fig:CA30}, we have depicted
rule 30 with each transition illustrated by three adjacent squares
representing the different preimages of $f$ and on the bottom, their
image by $f$. A 0 (resp. 1) is painted white (resp. black). On the
bottom of Fig.~\ref{fig:CA30}, we see the time-space diagram of the
cellular automaton from the \emph{initial configuration} at time $t=0$
to time $t=7$.

\section{LFSR synthesis by HCA}
\label{sec:synth}

We will restrict ourselves to the case where $Q=\mathbb{F}_2$ and $f$
is a Boolean predicate with 3 variables, an
\emph{elementary rule}. These CA have been considered
in~\cite{wolfram}: there are 256 different binary CA
and a natural number can be associated to each rule as follows:
\[\begin{array}{c|cccccccc}
x_{i-1}^t x_{i}^t x_{i+1}^t&111 &110 &101 &100 &011 &010 &001
&000\\\hline
x_i^{t+1}&0 &0 &0 &1 &1 &1 &1 &0 
\end{array}\]
The top line gives all possible preimages for $f$ and the bottom
line the images by $f$. Thus, $f$ is
fully specified by the 8-bit number written on the bottom line
(00011110 in our example) which can be translated in basis 10 and then
called the \emph{rule} of the cellular automaton (as rule
number 30 here).  Equivalently, this rule can be considered
as a Boolean function with (at most) 3 variables. Taking rule 30
again, its corresponding Boolean function is:
$x_i^{t+1}=x_{i-1}^t\oplus (x_{i}^t\vee x_{i+1}^t)$ with $\oplus$
denoting the Boolean XOR function and $\vee$ the classical Boolean OR
function. Its equivalent formulation in $\mathbb{F}_2$
is:
$x_i^{t+1}=
\left( x_{i-1}^t+x_{i}^t+x_{i+1}^t+x_{i}^tx_{i+1}^t\right)$.

\vspace*{-1cm}
\subsubsection{Equivalent rules}

Since we are dealing with pseudo-random generators, some of the
elementary rules are equivalent by three transformations, all
introduced by Wolfram in~\cite[p. 492]{wolfram}.  We first introduce
some notation: let us denote by $\tilde{w}$ the mirror image of the
finite binary word $w=w_1\ldots w_n$, $\tilde{w}=w_n\ldots w_1$ and by
$\overline{w}$ the word obtained from $w$ by exchanging the 0's by 1's
(and conversely) $\overline{w}=\overline{w_1}\ldots\overline{w_n}$.
The first transformation is the \emph{conjugation} which interchanges
the roles of 0 and 1. It takes as an input $r$, the binary
representation of a rule and returns $\overline{\tilde{r}}$. For
instance, the conjugation transforms rule 30 into rule 135.  The
second transformation, called \emph{reflection} gives a re-ordering of
the bits of $r$. Each bit $f_r(x_{i-1},x_i,x_{i+1})$ is replaced by
the value of $f_r(x_{i+1},x_i,x_{i-1})$ (the mirror image of
$x_{i-1}x_ix_{i+1}$) and leads to a re-ordering of the bits of $r$,
the binary representation of the rule. As an example, by reflection,
rule 30 is changed into rule 86. The last transformation combines
boths and is called \emph{conjugation-reflection}; it changes rule 30
into rule 149. All of these transformations keep the Walsh-Hadamard
transform values of the cellular automata dynamics and are thus
statistically equivalent.

\subsection{Hybrid Cellular Automata}

In the sequel, we will consider the case where different cells of the
CA can use different rules. This model is called hybrid and will be
denoted by HCA for short. In the
context of sequence generation, several authors have considered this
extension of the model of
CA~\cite{Cattell:1996aa,1016320,sipper96coevolving}. We will
focus on \emph{linear HCA} (LHCA) which is
used by \cite{Cattell:1996aa}.

\subsubsection{Linear hybrid cellular automata}

In \cite{SSMM,Cattell:1996aa}, Muzio et al. consider null-boundary hybrid
CA which only use two rules: rule 90 and 150. In this case, a CA is
fully specified by which cells use rule 90 and which use rule
150. This information is summarized in the \emph{rule vector}
$M=[d_0,d_1,\ldots,d_{N-1}]$ such that $d_i=\left\{
  \begin{array}[h]{ll}
    0&\mbox{ if cell $i$ uses rule 90}\\
    1&\mbox{ if cell $i$ uses rule 150}
  \end{array}
\right.$. Given $M$, its \emph{reversal} is $M$'s mirror image:
$[d_{N-1},\ldots,d_1,d_0]$. We also define the \emph{subvector}
$M_{i,j}=[d_i,\ldots,d_j]$ with $i\leq j$ which also represents a
submachine of the HCA consisting of cells $i$ through $j$.

The encoding of rules 90 and 150 into zero and one, resp., means that
equation~(\ref{eq:CArule}) can be rewritten in $\mathbb{F}_2$ as
$x_i^{t+1}=f_i(x_{i-1}^t,x_{i}^t,x_{i+1}^t)=x_{i-1}^t+d_ix_{i}^t+x_{i+1}^t.$
We define the \emph{state} of a HCA at time $t$ to be the $n$-tuple
formed from the state of the cells: $x^t=[x_0^t,x_1^t,\ldots,
x_{N-1}^t]^T$ (the superscript $^T$ denotes the transpose). Then, the
next state function of the HCA is computed as $x^{t+1}=f(x^t)$. Since
each $f_i$ is linear, $f$ is also linear and an endomorphism of
$\mathbb{F}_2^N$. Linearity implies the existence of a matrix $A$ such
that $x^{t+1}=f(x^t)=A\cdot x^t$. The \emph{HCA} transition matrix
plays the same role as an LFSR transition matrix. $A$ is tridiagonal.
\[A=\left(\begin{smallmatrix}
d_0 &1  &0 &\cdots &\cdots &0 &0\\
1   &d_1 &1 &\ddots &&&0\\
0&1&d_2&\ddots&\ddots&&\vdots\\
\vdots&&&&1&d_{N-2}&1\\
0&0&\cdots&\cdots&0&1&d_{N-1}\\
\end{smallmatrix}\right)\]
Let us denote by $\Delta$ the characteristic polynomial of $A$, that
is $\Delta=\mid x\mbox{Id}-A\mid$.
\begin{definition}\cite{Cattell:1996aa}
  A polynomial $p$ is said to be a HCA polynomial if it is the
  characteristic polynomial of some HCA.
\end{definition}
Recall that $M_{i,j}$ is the HCA consisting of cells $i$ through $j$
and denote $\Delta_{i,j}$ its corresponding characteristic
polynomial. When $i=0$, we simply write $M_k$
(resp. $\Delta_k$) for the CA consisting of cells $0$ to $k$
(resp. its corresponding characteristic polynomial). Cattell and
Muzio~\cite{Cattell:1996aa} proved that $\Delta_k$ satisfies a
recurrence relation:
\begin{theorem}\cite{Cattell:1996aa}\label{th:charpoly}
  $\Delta_k$ satisfies the reccurrence: $\Delta_{-2}=0,
  \Delta_{-1}=1$, $\Delta_k=(x+d_k)\Delta_{k-1}+\Delta_{k-2}$ for $k\geq 0$.
\end{theorem}
Theorem~\ref{th:charpoly} provides an efficient algorithm to compute
$\Delta_{N-1}$ the characteristic polynomial of a HCA from its rule
vector $M$. Actually, this recurrence relation is related to Euclidean
GCD algorithm on polynomials with $\Delta_k$ as the
dividend, $\Delta_{k-1}$ as the divisor, $x+d_k$ as the quotient and
$\Delta_{k-2}$ as the remainder. Applying Euclid's extended greatest
division algorithm yields to the sequence of quotients whose constant
terms are the mirror image of the rule vector. This comes
from:
\begin{lemma}\cite{Cattell:1996aa}\label{lem:HCAgcd}
  Let $p\in\mathbb{F}_2[x]$ and $q\in\mathbb{F}_2[x]$ of respective
  degrees $n$ and $n-1$. Then there exists a HCA with characteristic
  polynomial $p$ and characteristic subpolynomial $q$ if and only if
  applying Euclid's greatest division algorithm to $p$ and $q$ results
  in $n$ degree one quotients.
\end{lemma}
Thus $\Delta_{N-1}$ and $\Delta_{N-2}$ determine the whole HCA. But in
general, a characteristic polynomial isn't sufficient to uniquely
determine the HCA. Just consider the following counter-example:
$[0,0,1,0,0,0]\leftrightarrow x^6+x^5+x^4+x^3+1\leftrightarrow
[1,1,0,1,1,1]$

To uniquely determine the HCA, we must know one more characteristic
subpolynomial $\Delta_{1,N-1}$ and use theorem~\ref{th:CAquad}:
\begin{theorem}[HCA quadratic congruence~\cite{Cattell:1996aa}]\label{th:CAquad}
  Suppose we have a HCA with characteristic polynomial $\Delta_{N-1}$
  and characteristic subpolynomials $\Delta_{N-2}$ and
  $\Delta_{1,N-1}$. Then both $y=\Delta_{N-2}$ and $y=\Delta_{1,N-1}$
  satisfy the congruence:
  $y^2+(x^2+x)\Delta'_{N-1}y+1\equiv 0\mod \Delta_{N-1}$
  where $\Delta'_{N-1}$ is the formal derivative of $\Delta_{N-1}$ in
  $\mathbb{F}_2$.
\end{theorem}
By combining Lemma~\ref{lem:HCAgcd} and Theorem~\ref{th:CAquad},
Cattell and Muzio give a characterization of HCA polynomials and give
an algorithm for finding a HCA given a polynomial. Their method has
been recently improved in~\cite{LHCAsynt2:2007p63}.
\begin{corollary}
  Let $p\in\mathbb{F}_2[x]$ of degree $n$. Then $p$ is a HCA
  polynomial if and only if for some solution $q$ for y of the
  congruence
  \begin{equation}
    \label{eq:corCAquad}
    y^2+(x^2+x)p'y+1\equiv 0\mod p
  \end{equation} Euclid's greatest division
  algorithm on $p$ and $q$ results in $n$ degree one quotients.
\end{corollary}
Theorem~\ref{th:CAquad} has some weaknesses: it does not say neither
that polynomials solutions to the quadratic congruence will be
subpolynomials of $\Delta_{N-1}$ nor that non HCA polynomials
won't have solutions to the quadratic
congruence. Theorem~\ref{th:CAquad} only gives necessary conditions
for HCA polynomials: they have solutions to the quadratic congruence
and that some of these solutions are subpolynomials. However,
Theorem~\ref{th:CAquad} is useful for irreducible polynomials:
\begin{theorem}
  If $p\in\mathbb{F}_2[x]$ is an irreducible polynomial of degree $n$,
  then equation~(\ref{eq:corCAquad}) has exactly two solutions, both
  of which result in $n$ degree one quotients.
\end{theorem}
\begin{corollary}
  If $p\in\mathbb{F}_2[x]$ is an irreducible polynomial, then $p$ has
  exactly two HCA realizations with one being the reversal of the
  other.
\end{corollary}

Since one can build a HCA from an irreducible polynomial and represent
it by its transition matrix, we can ask which is the relationship
between LHCA and LFSR. If both are based on the same irreducible or
primitive polynomial, they have the same behavior up to permutation of
the order in which the states appear and the cycle structure of the
states is identical. A similarity transform between LHCA and LFSR has
been given in~\cite{Cattell:1998p61} and recently improved
in~\cite{kagaris:2006p60}.

\section{PRS generation by CA}
\label{sec:classical-PRS}

In \cite{wo-crypto,wo:alea}, Wolfram uses a
one-dimensional cellular automaton for pseudo-random bit generation by
selecting the values taken by a single cell when iterating the
computation of rule 30 from an initial finite configuration where the
cells are arranged on a ring of $N$ cells.  Mathematically, Wolfram
claims the sequence $\{x_i^t\}_{t\geq 0}$ is pseudo-random for a given
$i$. Wolfram extensively studied this particular rule, demonstrating
its suitability as a high performance randomizer which can be
efficiently implemented in parallel; indeed, this is one of the
pseudo-random generators which was shipped with the connection machine
CM2 and which is currently used in the Mathematica\textregistered\ software.

Unfortunately, this PRG is not suitable for cryptographic
purpose. In~\cite{ms91}, Meier and Staffelbach proposed a correlation
attack to reverse the PRS generated by rule 30 although it passes
classical statistical tests like the ones proposed in~\cite{knut}.

More recently, in~\cite{JCA-hiroshima}, we have used a Walsh transform
to explore the set of the 256 elementary rules.  The Walsh transform
is a well-known tool in the field of cryptology for studying the
correlation-immunity of Boolean functions: Xiao and
Massey~\cite{XiaoM88} have characterized the notion
of correlation-immunity with the Walsh transform. We have applied this
technique to the pseudo-random sequences generated by all of the 256
binary rules and we provide evidence that there does not exist a
non-linear rule which generates a correlation-immune pseudo-random
sequence. Thus, we state Theorem~\ref{th:pasgpa}.
\begin{theorem}\cite{JCA-hiroshima}\label{th:pasgpa}
  There is no non-linear correlation-immune elementary CA.
\end{theorem}

And, according to Theorem~\ref{th:yao}, we can state that:
\begin{corollary}
  There is no elementary CA which can serve as
  PRS generator.
\end{corollary}

So, does Theorem~\ref{th:pasgpa} annihilate any hope to design a good
PRG by the means of CA? Not
necessarily. Next section recalls the approach initiated by Tomassini
and Sipper and section~\ref{sec:appli} describe another way of
generating PRS with LHCA.

\section{PRS generation by HCA}
\label{sec:PRS-HCA}

\subsection{The cellular programming approach}

Tomassini and Sipper \cite{sipper96coevolving} proposed to use HCA for
generating better PRS. In this model, the rules are obtained by an
evolutionary approach (a genetic algorithm). They have designed a
\emph{cellular programming} algorithm for cellular automata to perform
computations, and have applied it to the evolution of pseudo-random
sequence generators. Their genetic algorithm uses Koza's
\emph{entropy} $E_h=-\sum_{j=1}^{k^h}p_{h_j}\log_2 p_{h_j}$ where $k$
denotes the number of possible values per sequence position, $h$ a
subsequence length and $p_{h_j}$ is a measured probability of
occurrence of a sequence $h_j$ in a pseudo-random sequence. It
measures the entropy for the set of $k^h$ probabilities of the $k^h$
possible subsequences of length~$h$. The entropy achieves its maximal
value $E_h=h$ when the probabilities of the $k^h$ possible sequences
of length $h$ are all equal to $1/\ell^h$, where $\ell^h$ denotes a
number of possible states of each sequence.
They have selected four rules of radius 1 for use in
non-uniform cellular automata. The best rules selected by the genetic
algorithm were rules $90$, $105$, $150$ and $165$ (which are all
linear, a clear drawback).

A series of tests (including $\chi^2$ test, serial correlation
coefficient, entropy and Monte Carlo, but no correlation-immunity
analysis) were made with good results, showing that co-evolving
generators are at least as good as the best available CA
randomizer. The authors also use elementary rules which we proved to
be not correlation-immune.  This was further investigated
in~\cite{1016320}.

\medskip

Following the same kind of approach, Seredynski et
al. in~\cite{1016320} have generalized the selection process to radius
2 rules. They use then both radius 1 and radius 2 rules in hybrid
cellular automata. The rules selected by their genetic algorithm were
$30$, $86$, $101$ and $869020563$, $1047380370$, $1436194405$,
$1436965290$, $1705400746$, $1815843780$, $2084275140$ and
$2592765285$.

Their new set of rules was tested by a number of statistical
tests required by the FIPS 140-2 standard~\cite{FIPS} but no
correlation-immunity analysis was made either.

\subsection{The synthesis approach}

This approach follows the synthesis algorithm proposed
in~\cite{Cattell:1996aa}. They propose a method for the synthesis of a
HCA from a given irreducible polynomial over $\mathbb{F}_2$. The same
problem for LFSR is well known as it can be directly obtained from the
transition matrix. Furthermore, there is a one to one correspondence
between LFSR's and polynomials. For CA, in general, a characteristic
polynomial is not sufficient to uniquely determine the CA from which
it was computed.

If we consider the characteristic polynomial $\Delta$ of the HCA
(assumed to be irreducible), with $\alpha$ a root in
$\mathbb{F}_{2^n}$. All $n$ roots of $\Delta$ lie in
$\mathbb{F}_{2^n}$. The roots $\alpha,\alpha^2,\alpha^{2^2},\ldots
,\alpha^{2^{n-1}}$ are distinct and $\Delta$ can be factored in
$\mathbb{F}_{2^n}$ as $(x-\alpha)(x-\alpha^2)(x-\alpha^{2^2})\ldots
(x-\alpha^{2^{n-1}})$.

\subsubsection{Product of irreducibles}

Given $p$ and $q$ two irreducible characteristic polynomials and $P$
and $Q$ their respective transition matrix, on can build the
transition matrix corresponding to $p\cdot q$. It can be defined by
blocks as:
$\left(\begin{smallmatrix}
  P&0\\
  0&Q
\end{smallmatrix}\right)$.
This operation corresponds to the concatenation of
LHCA~\cite{Cattell:1998p62}. They quoted that it permits to
concatenate primitive machines for forming machines of much longer
lengths.

\section{Application: boolean functions evaluation}
\label{sec:appli}

There is a well-known dictionary between, on the one hand, boolean
functions in $n$ variables, and binary sequences of period $2^n$. More
specifically, if $f$ is such a function, and if we denote by
$\overline{i}$ the base $2$ expansion of $i$ we can define a sequence
by the rule $s_f(i)=f(\overline{i}), $ for $i\le 2^n-1.$

Many interesting boolean functions can be cast under the form
$f(x)=Tr(ax+bx^s),$
where $a,b$ are scalars of the extension field $\mathbb{F}_{2^n}$ and
$Tr$ the trace function from $\mathbb{F}_{2^n}$ down to
$\mathbb{F}_{2}$. In the case where $n$ is odd and the Walsh Hadamard
transform takes only three values they are the so-called {\em
  plateaued} boolean function of order $n-1$~\cite{Z2} also known as
almost optimal or semi-bent.

They are the traces of so-called {\em almost bent} $AB$
functions~\cite{CCZ-98}.
For monomials $AB$ functions, the most famous exponents $s$ are in the
conjecturally exhaustive list of Gold, Kasami, Welch, Niho (see
Table~\ref{tab:s-power}).
\vspace*{-.3cm}
\begin{table}[h]
  \centering
  \begin{tabular}{|c|c|c|}\hline
    Name&$s$&Condition\\\hline\hline
    Gold&$2^i+1$&$i\wedge m=1, 1\leq i\leq m/2$\\
    Kasami&$2^{2i}-2^i+1$&$i\wedge m=1, 1\leq i\leq m/2$\\
    Welch&$2^{(m-1)/2}+3$&\\
    Niho&$2^{2r}+2^r-1$&
    $\begin{array}[t]{c}
      r=t/2\mbox{ for $t$ even}\\
      r=(3t+1)/2\mbox{ for $t$ odd}\\
      \mbox{with $1\leq r\leq m=2t+1$}
    \end{array}$\\\hline
  \end{tabular}
  \caption{\sf{Exponents of AB monomials.}}
  \label{tab:s-power}
\end{table}
\vspace*{-1cm}
In all these cases, an upshot of the theory of Mattson-Solomon
polynomials \cite[p.249]{macsloane} is that the parity check
polynomials of the attached cyclic codes (or, essentially, the
connection polynomial of the LFSR) is of the form
$m_{\alpha}m_{\alpha^s},$ where $\alpha$ generates $\mathbb{F}_{2^n}$
over $\mathbb{F}_{2}$. A fast algorithm to compute the minimal
polynomials of elements in finite field extensions is given
in~\cite{AC}.

\section{Conclusion}
We have used the synthesis approach to give an effective
CA-realization of classical pseudo-random sequences of cryptographic
quality. The main interest of this work would be to give an hardware
implementation. The target hardware model of CAs is the Field
Programmable Gate Arrays (known as FPGAs). FPGAs are now a popular
implementation style for digital logic systems and subsystems. These
devices consist of an array of uncommitted logic gates whose function
and interconnection is determined by downloading information to the
device. When the programming configuration is held in static RAM, the
logic function implemented by those FPGAs can be dynamically
reconfigured in fractions of a second by rewriting the configuration
memory contents. Thus, the use of FPGAs can speed up the computation
done by the cellular automata.  Putting all together allows high-rate
pseudo-random generation of good quality.

\bibliographystyle{plain}
\bibliography{toute}
\end{document}